\documentclass[twocolumn,10pt]{article} 

\usepackage[square,numbers,sort&compress,comma]{natbib}

\usepackage{amsmath}
\usepackage{amssymb}
\usepackage{caption}
\usepackage{graphicx}
\usepackage{latexsym}
\usepackage{times,hyperref}
\usepackage[pagewise]{lineno}
\usepackage{multirow}

\topmargin - 12pt 
\renewenvironment{abstract}%
              {
               \small
               {\bfseries \abstractname}
               \par
               \vspace{10pt}
              }

\renewcommand\abstractname{Abstract}

\newcommand{\nomenclature}
              [1]
              {
               \bgroup
               \flushleft
               \small\bf
               #1
               \par
               \egroup
              }

\renewcommand{\section}
              [1]
              {
               \bgroup
               \flushleft
               \small\bf
               \refstepcounter{section}
               \arabic{section}. #1
               \par
               \egroup
              }

\renewcommand{\subsection}
              [1]
              {
               \bgroup
               \flushleft
               \small\em
               \refstepcounter{subsection}
               \arabic{section}.
               \arabic{subsection}. #1
               \par
               \egroup
              }

\renewcommand{\subsubsection}
              [1]
              {
               \bgroup
               \flushleft
               \small\em
               \refstepcounter{subsubsection}
               \arabic{section}.
               \arabic{subsection}.
               \arabic{subsubsection}. #1
               \par
               \egroup
              }

  \newcommand{\acknowledgement}
              [1]
              {
               \bgroup
               \flushleft
               \small\bf
               #1
               \par
               \egroup
              }

  \newcommand{\sectionbib}
              [1]
              {
               \bgroup
               \flushleft
               \small\bf
               #1
               \par
               \egroup
              }

\begin{document}

\title{\LARGE  Detailed simulation of LOX/GCH4 flame-vortex interaction in supercritical Taylor-Green flows with machine learning\\
                    }

\author{{\large Jiayang Xu$^{b,\#}$, Yifan Xu$^{a,b,\#}$, Zifeng Weng$^{c,b}$, Yuqing Cai$^{d,b}$, Runze Mao$^{a,b}$,}\\[10pt]
{\large  Ruixin Yang$^{a,b}$, Zhi X. Chen$^{a,b,*}$}\\[10pt]
        {\footnotesize \em $^a$State Key Laboratory of Turbulence and Complex Systems, Aeronautics and Astronautics, College of Engineering,}\\[-5pt]
        {\footnotesize \em Peking University, Beijing, 100871, PR China}\\[-5pt]
        {\footnotesize \em $^b$AI for Science Institute (AISI), Beijing, 100080, PR China}\\[-5pt]
        {\footnotesize \em $^c$Center for Combustion Energy, School of Vehicle and Mobility}\\[-5pt]
        {\footnotesize \em Tsinghua University, Beijing, 100084, PR China}\\[-5pt]
        {\footnotesize \em $^d$State Key Laboratory of Engines}\\[-5pt]
        {\footnotesize \em Tianjin University, Tianjin 300072, PR China}
        }

\date{}


\small
\baselineskip 10pt


\twocolumn[\begin{@twocolumnfalse}
\vspace{50pt}
\maketitle
\vspace{40pt}
\rule{\textwidth}{0.5pt}
\begin{abstract} 
Accurate and affordable simulation of supercritical reacting flow is of practical importance for developing advanced engine systems for liquid rockets, heavy-duty powertrains, and next-generation gas turbines. In this work, we present detailed  numerical simulations of LOX/GCH4 flame-vortex interaction under supercritical conditions. The well-established benchmark configuration of three-dimensional Taylor-Green vortex (TGV) embedded with a diffusion flame is modified for real fluid simulations. Both ideal gas and Peng-Robinson (PR) cubic equation of states are studied to reveal the real fluid effects on the TGV evolution and flame-vortex interaction. The results show intensified flame stretching and quenching arising from the intrinsic large density gradients of real gases, as compared to that for the idea gases. Furthermore, to reduce the computational cost associated with real fluid thermophysical property calculations, a machine learning-based strategy utilising deep neural networks (DNNs) is developed and then assessed using the three-dimensional reactive TGV. Generally good prediction accuracy is achieved by the DNN, meanwhile providing a computational speed-up of 13 times over the convectional approach. The profound physics involved in flame-vortex interaction under supercritical conditions demonstrated by this study provides a benchmark for future related studies, and the machine learning modelling approach proposed is promising for practical high-fidelity simulation of supercritical combustion. 
\end{abstract}
\vspace{10pt}

\parbox{1.0\textwidth}{\footnotesize {\em Keywords:} Flame-vortex interaction; Supercritical combustion; Real fluid; Machine learning; Taylor-Green Vortex}
\rule{\textwidth}{0.5pt}
\vspace{10pt}
*Corresponding author.  $^{\#}$ These two authors contribute equally to this work.\\
\textit{E-mail address:} chenzhi@pku.edu.cn (Zhi X. Chen).\\
\vspace{10pt}
\end{@twocolumnfalse}] 

\clearpage

\twocolumn[\begin{@twocolumnfalse}

\centerline{\bf Information for Colloquium Chairs and Cochairs, Editors, and Reviewers}

\vspace{20pt}



\vspace{20pt}

{\bf 1) Novelty and Significance Statement}


\vspace{10pt}

The novelty of this research is the establishment of a new supercritical, real-gas combustion benchmark with TGV for vortex-flame interaction, and the application of machine-learning to improve the calculation efficiency without scarifying the accuracy. It is significant because rare three-dimensional benchmark exits for investigating the vortex-flame interaction under supercritical conditions. To the best knowledge of the authors, this work is the first attempt to establish such a benchmark, giving insight to the community regarding the complex vortex-flame interaction phenomenon in the context of diffusion flame under supercritical conditions.

\vspace{20pt} 

{\bf 2) Author Contributions}
\vspace{10pt}


\begin{itemize}

  \item{Jiayang Xu's contributions: developed code, designed research, performed research, analyzed data, wrote the paper }

  \item{Yifan Xu's contributions: analyzed data, wrote the paper}
  
  \item{Zifeng Weng's contributions: wrote code, implement APIs for NNs}
  
  \item{Yuqing Cai's contributions: literature review}

  \item{Runze Mao's contributions:wrote code} 

  \item{Ruixin Yang's contributions:wrote code}

  \item{Zhi X. Chen's contributions: supervision, reviewed and edited paper, funding acquisition.}

\end{itemize}

\vspace{10pt}

{\bf 3) Authors' Preference and Justification for Mode of Presentation at the Symposium}
\vspace{10pt} 


The authors prefer {\bf OPP} presentation at the Symposium, for the following reasons:

\begin{itemize}

  \item{The application of supercritical combustion with well-established Taylor-Green Vortex is a pioneering attempt.}

  \item{Some flame-vortex interaction behaviours revealed potentially can contribute to future research’s validation. }

  \item{Conclusions can be drawn through meticulously depicted graphs of detailed physical structures without extensive background information.}

\end{itemize}

\end{@twocolumnfalse}] 


\clearpage


\section{Introduction\label{sec:introduction}} \addvspace{10pt}

Transcritical and supercritical reacting flow plays an important role in the development of liquid rocket engines (LREs), as combustion chambers of LREs are designed to operate under pressures well above the critical points of injected fuels and oxidizers to produce more specific impulse and thrust~\cite{YAO2019302,SINGLA20052921}. Understanding ultra-high pressure fluid and combustion behaviour is of high practical importance for developing advanced space propulsion devices~\cite{YANG2000925}. 

Under these extreme pressures, as the interactions between molecules cannot be neglected, the assumption of perfect gas no longer holds and real gas effects become significant~\cite{LACAZE20151603}. The thermophysical properties of real gas exhibit large gradients in the temperature-pressure phase diagram~\cite{inbook}. Due to the simplicity and easy implementation, cubic equations such as Peng-Robinson (PR)~\cite{PR} and Soave-Redlich-Kwong(SRK)~\cite{SRK} are extensively used as real-gas equation of state (EOS) to fit the variations of transport and thermodynamic properties. For a broad range of interest, satisfactory accuracy is obtained when comparing these EOSs with the NIST database~\cite{NIST}. However, because of the high non-linearity of real-gas thermophysical properties, severe nonphysical numerical oscillations have been reported in many studies~\cite{MA2017330}. Another obstacle for real fluid simulation is the formidable computational cost. Under trans/supercritical conditions, resolving the large scalar gradients requires highly refined grids, limiting the computational efforts to two-dimensional or specific domain and geometries~\cite{bellen}. 

Flame-vortex interaction is of practical importance in many combustion systems including LREs and other propulsion devices~\cite{RENARD2000225}. The strong shear flows with large density gradient present in LREs further enhances the dynamical behaviour of flame-vortex interactions. The Taylor-Green vortex~\cite{Taylor} is a well-defined and broadly employed configuration to study vortex dynamics. It is also regarded as benchmark case for computational fluid dynamics (CFD)~\cite{WANG} and combustion \cite{ABDELSAMIE2021104935} code validation.  
Recently, a number of combustion modelling groups have attempted to simulate the TGV-flame interaction, not only for code accuracy and efficiency assessment~\cite{ABDELSAMIE2021104935,ABDELSAMIE2016123,TGV} but also to study flame-vortex interaction under various conditions~\cite{Yifan}. In the context of supercritical fluids, a few benchmarks have also been established to cross validate different in-house codes. Ruiz et al. ~\cite{ruiz2016numerical} proposed a two-dimensional mixing layer with an injector lip separating liquid-oxygen (LOX) and gaseous-hydrogen (GH2) streams at high Reynolds numbers. Ma and co-workers~\cite{MA2017330} also developed various numerical methods based on this benchmark. Muller et al.~\cite{muller} simulated a configuration for coaxial injection of liquid-nitrogen (LN2) and GH2 using a pressure-based OpenFOAM solver and a density-based high-order solver. They found that the thermophysical properties presented by the resolved scales were more important than the subgrid closures and exact code used. However, the above progresses in developing a benchmark for supercritical fluid simulation were limited to two-dimensional and non-reactive conditions. 

With these missing pieces and gaps in the literature, the objective of this work is threefolds. First, it is of interest to establish a numerical benchmark for three-dimensional trans/supercritical reactive flows, which can be used for code verification and validation. Second, by combining the rich features of TGV-flame setup and typical real fluid conditions with large density gradients, interesting flame-vortex interaction behaviours can be revealed. Last but quite importantly, data-driven methods have facilitated new efficient modelling of the computationally expense real fluid properties calculation~\cite{MILAN2021110567,ZHOU2024124888}, and here we also present a modelling strategy to accelerate the heavy three-dimensional real fluid simulation with a machine learning approach.

The remainder of this paper is organised as follows. Section~\ref{sec:NM} introduces physical models and equations and the numerical tools used. Section~\ref{sec:TGV} demonstrates two-dimensional cold flow, three-dimensional mixing, and three-dimensional reacting TGV. Details of the DNNs and its performance when applied to the three-dimensional reacting case are discussed in Section~\ref{sec:DNN}. Conclusions are summarised in Section~\ref{conclusion}.


\begin{figure*}[h!]
\centering
\vspace{-0.6 in}
\includegraphics[width=363pt]{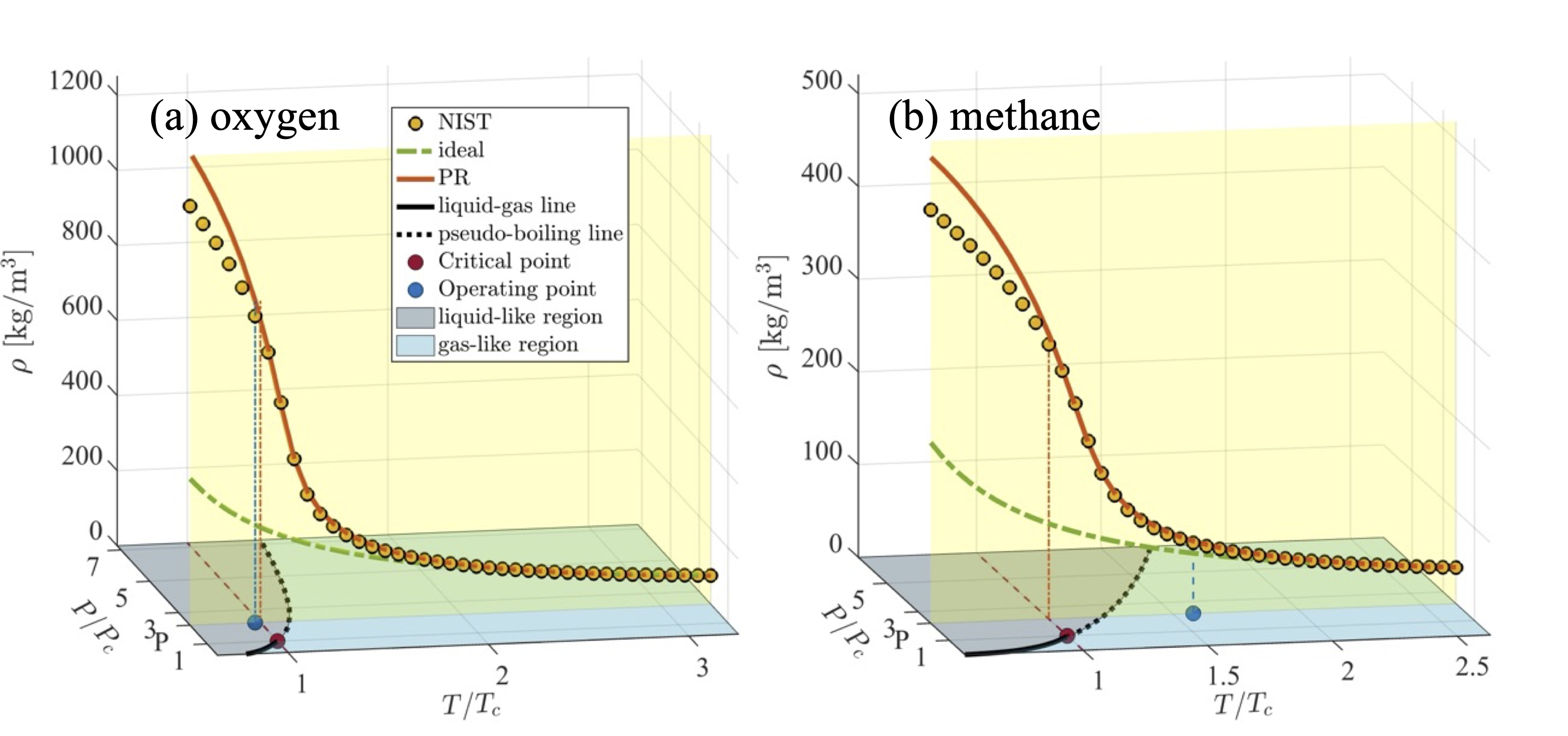}
\vspace{10 pt}
\caption{\footnotesize The comparison between states of oxygen (a) and methane (b) gained from ideal gas equation, PR model, and NIST database. The operating conditions for cases in Section~\ref{sec:TGV} are labelled by blue dots. }
\label{Trho}
\end{figure*}

\section{Numerical method\label{sec:NM}} \addvspace{10pt}
The numerical solver used in this study is a pressure-based weakly compressible low-Mach flow solver~\cite{df} developed from the OpenFOAM package. The PISO algorithm is modified to update thermodynamic properties every inner pressure correction loop for improved numerical stability~\cite{Jarczyk2012LargeES}. Detailed molecular transport is implemented via a code coupling with Cantera. The chemical kinetics are calculated using the CVODE method. 
The real fluid thermophysical properties are calculated using Cantera~\cite{Cantera} and the Peng-Robinson equation is used as the cubic EOS, which is written as 
\begin{equation}
p = \frac{RT}{V-b} - \frac{a}{V^2+2bV-b^2},
\end{equation}
where the coefficients $a$ and $b$ incorporate the effects of intermolecular forces, quantitatively represented as $a=0.45724\frac{R^2T_c^2}{P_{c}}$ and $b=0.07780\frac{RT_c}{P_{c}}$ for pure substances, where $T_c$ is the critical temperature and $P_c$ is the critical pressure. For mixtures, the PR equation is adapted using the Van der Waals' mixing rules to accommodate interactions between multiple species. 

The PR model is selected as the cubic EOS for its better performance for temperatures higher than the critical temperature $T_c$. As depicted in Fig.~\ref{Trho}, the cubic EOS displays a robust ability to model large density gradients for the oxygen and methane states considered. The PR model also exhibits good agreement with the NIST reference data for O$_2$ above its $T_c = 154$~K and for CH$_4$ above $T_c = 190$~K at a constant pressure of $10$~MPa. The model efficacy extends across a broad temperature range, effectively capturing the transition of fluids across the pseudo-boiling line from 'liquid-like' to 'gas-like' states. Takahashi's high-pressure correction~\cite{1975417} is used for binary diffusion coefficients, a modified Chung’s transport model~\cite{Chung} with mole-fraction-averaged viscosity is used for dynamic viscosity and thermal conductivity.

 \section{Supercritical TGV-flame configuration\label{sec:TGV}} \addvspace{10pt}
Following the previous benchmark study \cite{ABDELSAMIE2021104935}, the proposed supercritical TGV-flame setup is demonstrated using three steps: i) a two-dimensional cold flow, ii) a three-dimensional cold flow, and iii) a three-dimensional reacting flow. A cubic domain of $[0,L]^n$ is simulated, where $L = 2\pi L_0$ and n is the dimension. The reference time is defined by the vortex turn-over time:~$\tau_{ref} = {L_0}/{u_0} = 7.5$~$\mu$s, where $u_0$ is the initial velocity magnitude. Cases are designed to have the identical $L_0 = 0.03$~mm and $u_0 = 4$~m/s to achieve $Re = {u_0*L_0}/{\nu_{min}} \approx 1600$~\cite{ABDELSAMIE2021104935,TGV}. Further details including spatial resolutions are summarised in Table~\ref{cases}.

\begin{table}[h!] \footnotesize
\caption{Summary of cases}
\centerline{\begin{tabular}{c c c c}

\hline
Cases                 & $\nu_{min}$     & Reynolds & {resolution}  \\
 & ($10^{-8}$m$^2$/s) & number & (\# grids)\\
\hline
Step 1                       &   7.7548 & 1547    &  256 \\
Step 2                      &  7.7548   & 1547   & 512\\
Ideal gas                   & 6.5869   & 1821 & 256 \\ 
Real gas                     & 7.7548     & 1547 & 256 \\ 
\hline 
\end{tabular}
}

\label{cases}
\end{table}

\subsection{Step 1: Two-dimensional non-reacting flow\label{subsec:2D}} \addvspace{10pt}

\begin{figure}[h!]
\centering
\includegraphics[width=192pt]{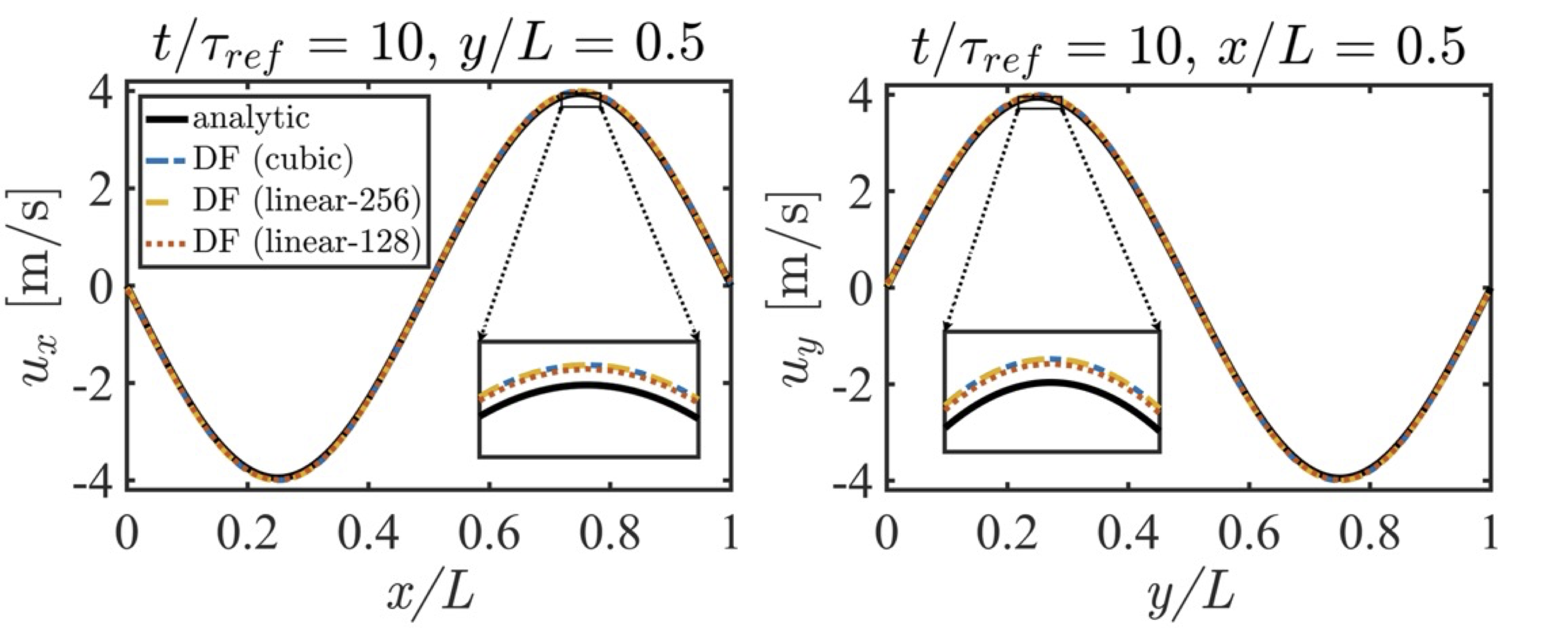}
\caption{\footnotesize Comparison of central line x-component (left) and y-component (right) velocity profile with different resolutions and spatial schemes.}
\label{Uxy}
\end{figure}

For two-dimensional TGV with a small Reynolds number, an analytical solution can be obtained. In this configuration, liquid oxygen at $150$~K and $10$~MPa is selected as the fluid with the kinematic viscosity of $\nu = 7.7548\times 10^{-8}$~m$^2$/s. The analytical solution for the velocity is given by
\begin{equation}
u(x,y,t)= u_0 \times sin(\frac{x}{L_0})\times cos(\frac{y}{L_0})\times e^{\frac{-2\nu t}{L_0^2}} ,
\end{equation}
\begin{equation}
v(x,y,t) = -u_0 \times cos(\frac{x}{L_0})\times sin(\frac{y}{L_0})\times e^{\frac{-2\nu t}{L_0^2}} .
\end{equation}Figure~\ref{Uxy} shows the comparison between the analytical solution and simulation results at $t= 10 \tau_{ref}$, using different grid resolutions and spatial schemes. The results suggest that the linear scheme (2nd order central differencing) provides a satisfactory accuracy as in the 4th order cubic scheme. Also, resolutions of $N = 128$ and $256$ give no obvious difference, indicating a good numerical grid convergence. 

\begin{figure}[h!]
\centering
\includegraphics[width=192pt]{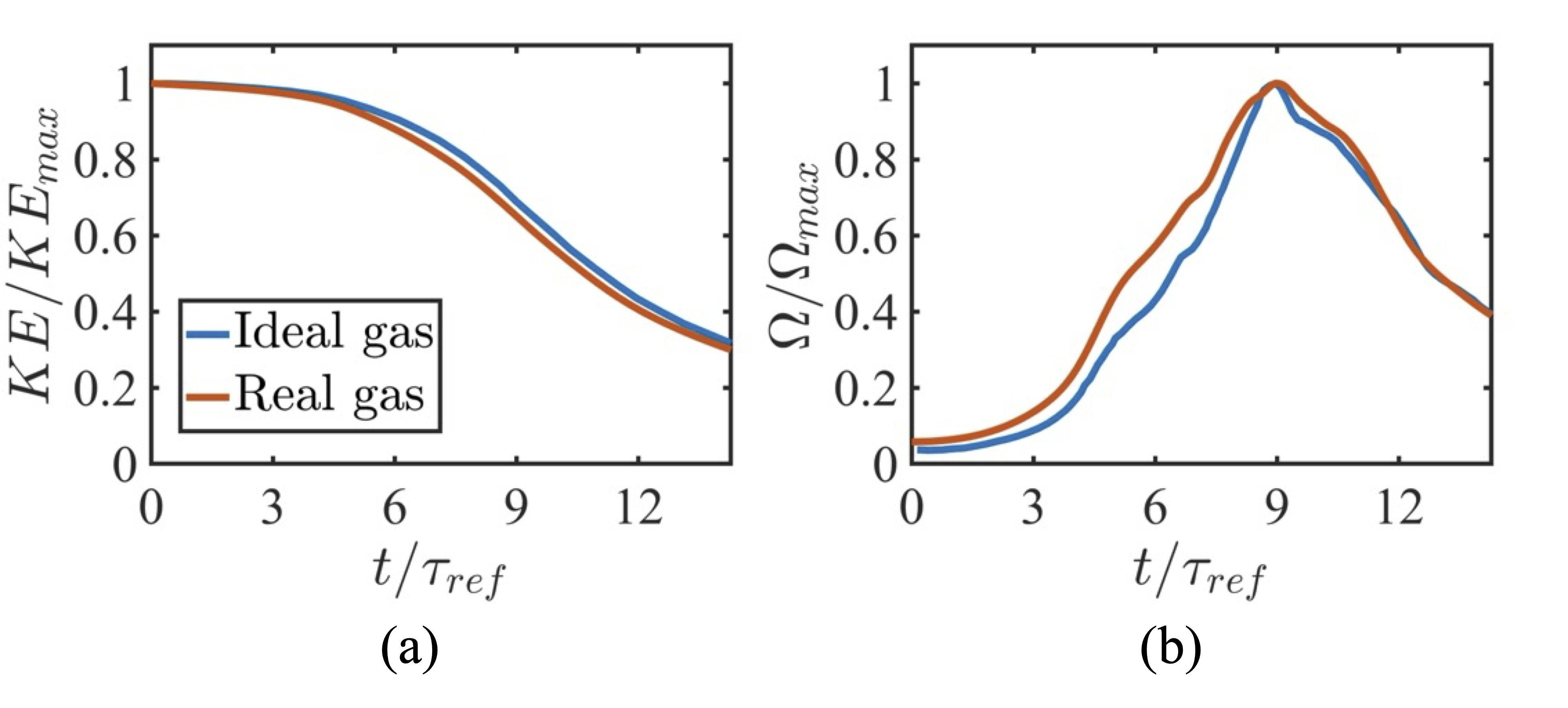}
\caption{\footnotesize Comparison of the temporal evolution of normalized turbulent kinetic energy (a) and of its dissipation rate (b), known as enstrophy, with ideal gas reference data from~\cite{VANREES20112794,ABDELSAMIE2021104935}.}
\label{step2}
\end{figure}

\subsection{Step 2: three-dimensional non-reacting flow\label{subsec:3D-cold}} \addvspace{10pt}
This step aims to further investigate the fluid processes in TGV with the real gas of oxygen at $150$~K and $10$~MPa. In Fig~\ref{step2}, the temporal evolution of the normalised volume-averaged kinetic energy and enstrophy is compared to the non-dimensional results obtained with a pseudo-spectral method in \cite{VANREES20112794,ABDELSAMIE2021104935}. The figure suggests that the progression of non-dimensional enstrophy under extreme pressure and thermodynamic conditions in cold flow scenarios exhibits a similar pattern to that in subcritical cases. A notable difference is that the enstrophy increases faster in the supercritical case, which is mainly due to the larger density gradients resulting in stronger vortex formation as observed in previous mixing layer studies~\cite{ruiz2016numerical}.

\begin{figure}[h!]
\centering
\includegraphics[width=192pt]{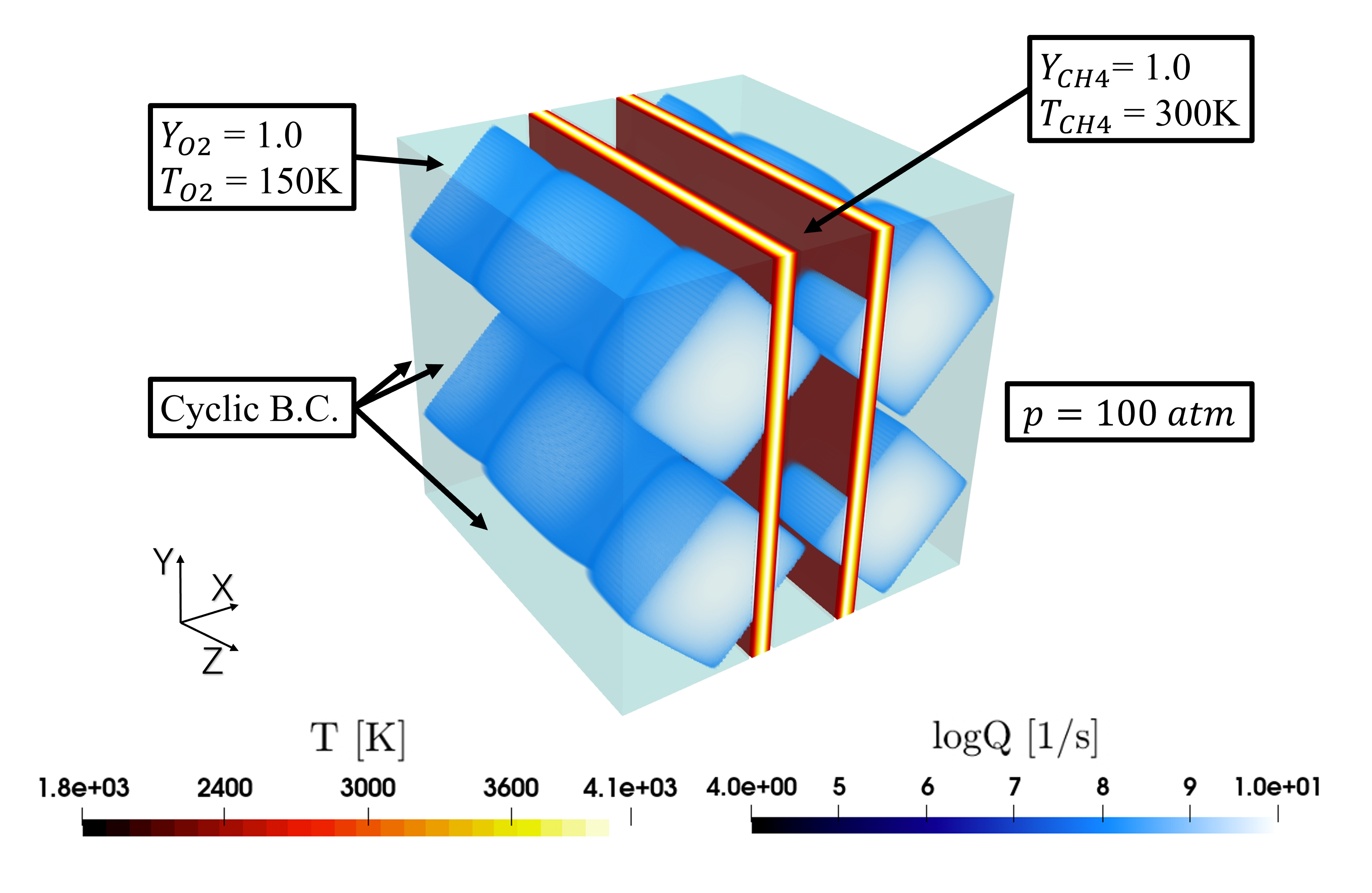}
\caption{\footnotesize The initial field of the 3D reacting TGV, with temperature iso-surface indicating flame location, log($Q$-criterion) indicating vortices. Initial temperature and species' mass fractions are labelled.}
\label{ini}
\end{figure}

\begin{figure*}[h!]
\centering
\vspace{-0.6 in}
\includegraphics[width=320pt]{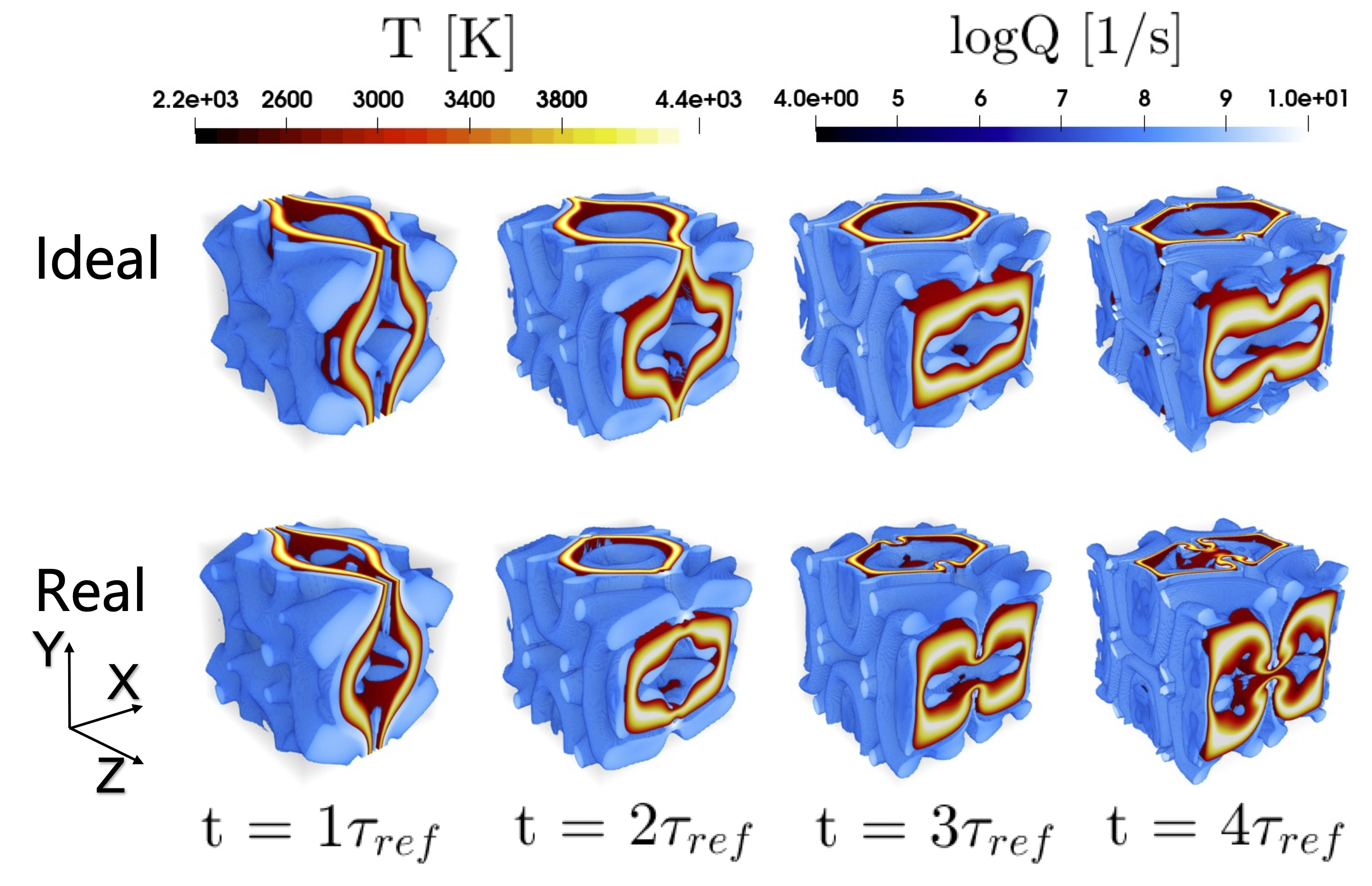}
\vspace{10 pt}
\caption{\footnotesize Evolution of the temperature field and logarithm of the $Q$-criterion field for the 2 cases. In
each line, the fields are displayed from $t = 1\tau_{ref}$ to $4\tau_{ref}$ for each case, where the same colormaps of temperature and log$Q$ are given accordingly.}
\label{4ref}
\end{figure*}

\begin{figure*}[h!]
\centering
\includegraphics[width=380pt]{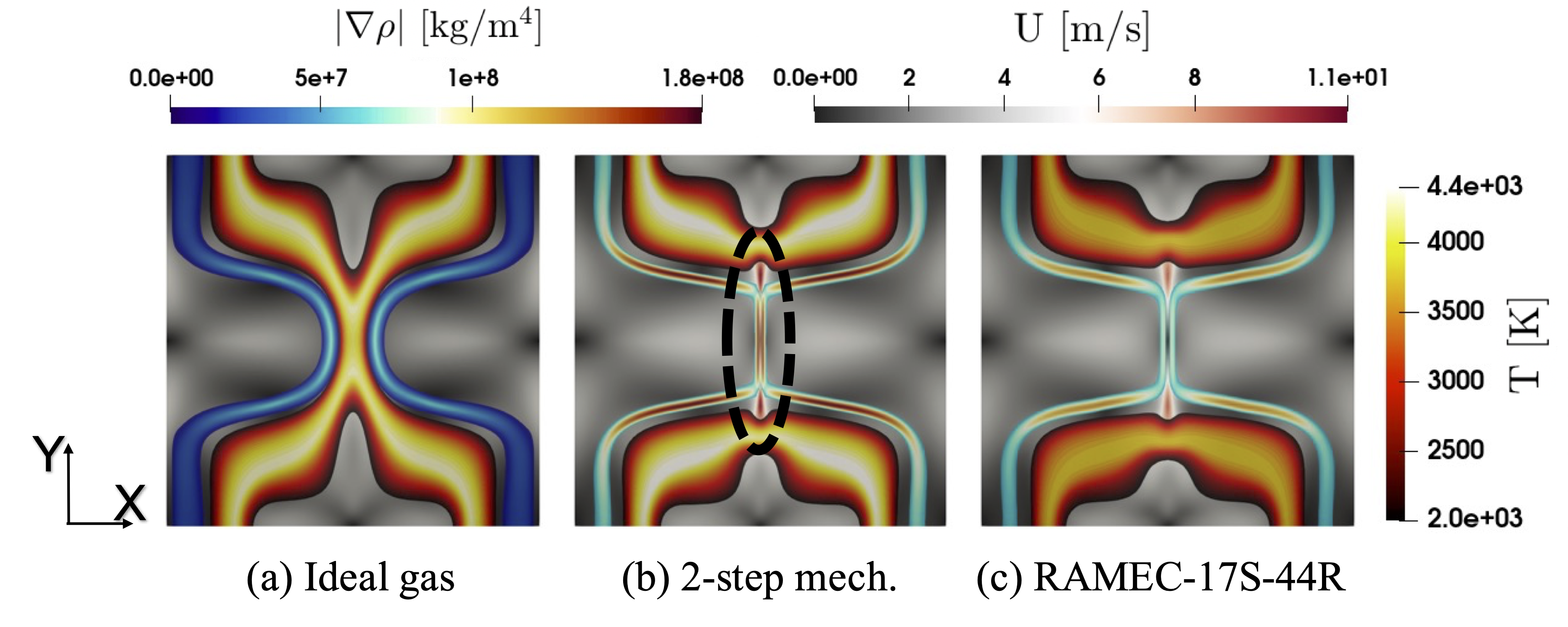}
\vspace{10 pt}
\caption{\footnotesize Comparison of $\nabla \rho$, temperature and velocity magnitude fields at x-y cross-section through the centre of the domain among the three cases: (a) ideal gas, (b) real gas using 2-step mech. and (c) real gas using the RAMEC-17S-44R mech.}
\label{2ref}
\end{figure*}

\subsection{Step 3: Three-dimensional reacting flow\label{subsec:3DreactiveTGV}} \addvspace{10pt}
Initial fields of temperature and $Q$-criterion for the three-dimensional multi-species reacting case are shown in Fig~\ref{ini}. This setup closely follows that of \cite{ABDELSAMIE2021104935} to allow for a reasonable comparison with the ideal gas benchmark. 
\begin{figure}[h!]
\centering
\includegraphics[width=192pt]{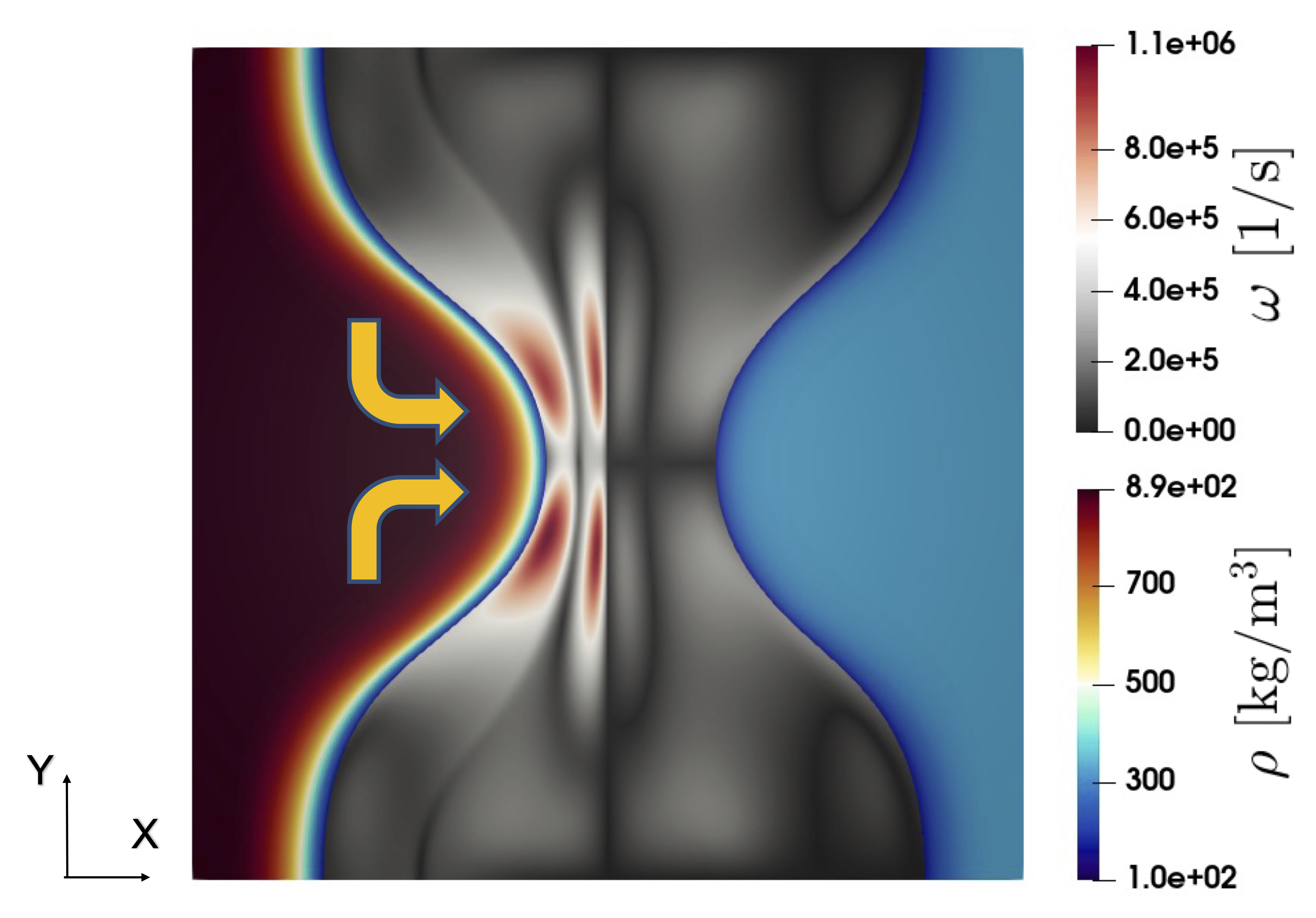}
\caption{\footnotesize Comparison of the vorticity and density fields of the real gas case (left half) and ideal gas case (right half) on the same graph of the x-y cross-section using the same colorbar.}
\label{1ref}
\end{figure}
The differences lie in the thermodynamic states, for which a supercritical pressure of $10$~MPa is specified uniformly across the domain. The initial temperature and species distributions are similar to those in \cite{ABDELSAMIE2021104935}, but for a CH$_4$/LO$_x$ mixture with unburnt oxygen at $150$~K and methane at $300$~K. 
The high temperature region is interpolated from the equilibrium state calculated from a two-step five-species ~\cite{FRANZELLI2012621} mechanism.
To compare the differences of the flame-vortex interactions using the ideal gas model and PR model, the isovolume of temperature and logarithm of $Q$ fields are depicted from $t = 1$ to $4\tau_{ref}$ for both cases in Fig~\ref{4ref}. The $Q$-criterion is defined as $Q = \frac{1}{2}~(\|\boldsymbol{\Omega}\|^2-\|\boldsymbol{S}\|^2)$, where $\boldsymbol{\Omega}$ is the vorticity tensor and $\boldsymbol{S}$ is the strain rate tensor. The positive values of $Q$ are indicative of areas in the flow field where the vorticity dominates and negative values of $Q$ are indicative of strain rate or viscous stress dominated areas. Thus, logarithm of $Q$ is used in Fig~\ref{4ref} to visualise vortical structures. It is seen that similar vortical structures can be observed in both cases, and the initial lumpy vortices are rolled up into vortex tubes surrounded by the spreading flame surface from $t = 1$ to $2\tau_{ref}$. Parts of the tubes are then stretched and twisted into thinner ones in following time instants. Considering the flame-vortex interaction, the ideal gas case has the identical flame structures and evolving stages as compared to the hydrogen TGV-flame at subcritical conditions simulated using DNS with a similar Reynolds number~\cite{Yifan}. At the same reference times, although the temperature under extreme pressure is much higher, the deformations of the flame and the stretching of the vortices are similar. By contrast, the interactions between the flame and vortex in real gas case start to behave differently from $t = 2\tau_{ref}$. Compared to the ideal gas case, more pronounced deformations of the flame surface occur due to stronger vortices and higher strain rate. Moreover, breaking and quenching of the flame are observed especially in the $y$-direction.

To further understand the differences appearing from $t= 2\tau_{ref}$, the contour plots for the $x$-$y$ cross-section plane is presented in Fig.~\ref{2ref}. Several evident differences can be observed in this figure. Firstly, the real gas case exhibits significant density gradient near the flame due to the substantial difference in density between the LOX and GCH4 mixtures. Moreover, there are two regions with notably high $U_y$ around $y$ central line (marked by black dashed lines), swirling and stretching the flame away from the centre of the domain. As seen in Fig.~\ref{2ref}b, flame quenching occurs in this region due to the high strain rates, which is significantly different to the ideal gas case in Fig.~\ref{2ref}a. To eliminate the possible influence of chemical mechanism, the simulation performed using a detailed 17-species/44-step mechanism~\cite{MONNIER20232747,MONNIER2022111735} is presented in Fig.~\ref{2ref}c (see detailed quantitative comparison in the Supplementary Material). Despite the considerably low peak temperature, the detailed chemistry case exhibit similar flow and flame patterns as the 2-step mechanism case. In particular, the flame quenching induced by the high strain near the central line is also observed, suggesting that this behaviour is mainly driven by the real gas thermodynamic effects. 


To investigate how these differences are generated, vorticity and density fields at $t = 1\tau_{ref}$ are compared between the real gas and ideal gas cases in Fig~\ref{1ref}. As mentioned in Section~\ref{sec:NM}, the unburned oxygen lies in the `liquid-like’ region while the high-temperature mixture are more `gas-like’ during the TGV evolution. Similar to that observed in~\cite{doi:10.2514/6.2013-3717}, the significant density gradients lead to a dramatic change in the momentum of the two regions, and thus the dense oxygen acts as a ‘solid obstacle’ against the rotating internal gaseous part of the flow. Comparing the two cases in this figure, it can be seen that due to the constraining effect of the 'solid obstacle', when the vortex rolls up the flame surface to the centre, the flame appears in a narrower region near the centre of the $x$-$y$ plane with higher vorticity. The stronger vorticity and strain rate in the real gas case leads to further stretching and quenching of the flame as shown earlier in Fig~\ref{2ref}. 

\begin{figure}[h!]
\centering
\includegraphics[width=192pt]{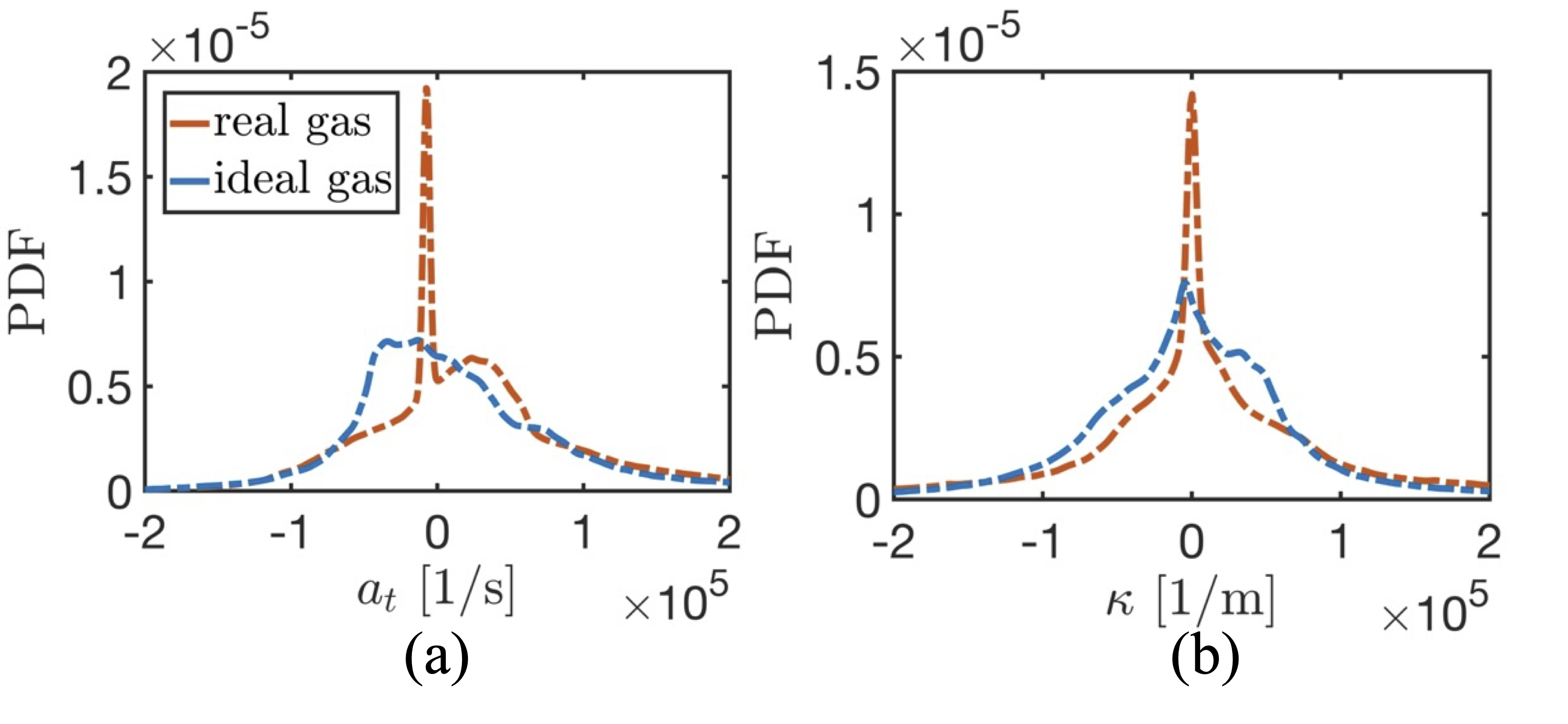}
\caption{\footnotesize The PDF of flame tangential strain rate (a) and curvature (b) of both cases at $t = 4\tau_{ref}$.}
\label{curvature}
\end{figure}

To shed further light into the flame-vortex interaction, Fig.~\ref{curvature} presents the Probability Density Function (PDF) of curvature and tangential stain rate for the two cases at $t = 4\tau_{ref}$. The positive tangential strain rate of real gas with higher PDF is seen in Fig.~\ref{curvature}a for the real gas case indicating a stronger flame stretching, which is consistent with the curvature PDF in Fig.~\ref{curvature}b. The high peak of curvature distribution around zero in the real gas case also implies that the flame is globally less deformed but with substantial local wrinkles (a second peak exists) compared to the more distributed shape in the ideal gas case. 


\begin{figure*}[h!]
\centering
\vspace{-0.6 in}
\includegraphics[width=363pt]{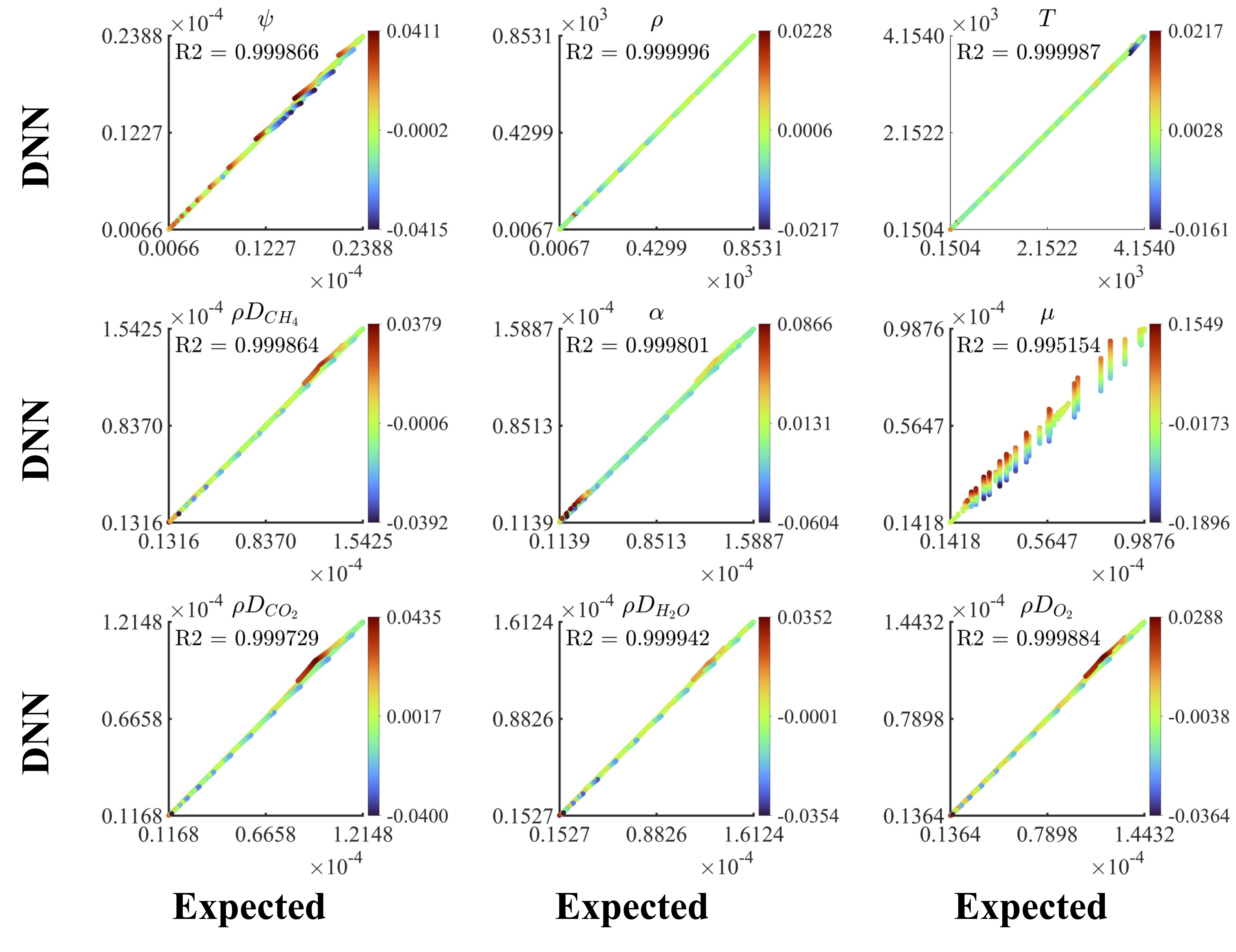}
\vspace{10 pt}
\caption{\footnotesize The regression plots of the DNN approximation values over expected values for 9 selected variables out of 17 in total, colored by local relative errors. All variables are expressed in SI units.}
\label{dnn}
\end{figure*}

\section{Machine learning modelling\label{sec:DNN}} \addvspace{10pt}
To simulate high-resolved three-dimensional supercritical reactive flows, the high computational cost remains a major obstacle. To facilitate broader application of such detailed simulations, a machine learning approach is proposed here to improve the efficiency of the most time-consuming real fluid thermophysical property computations. Deep neural networks are trained to learn the PR equation, mixing rules, and transport models. 

\subsection{Network architecture\label{subsec:NA}}
\addvspace{10pt}
The current investigation adopts a multilayer perceptron (MLP) framework, incorporating $3$ hidden layers with the Gaussian Error Linear Unit (GeLU) activation function. Inputs and outputs are tailored for the CFD solver used, where the real fluid properties as model outputs are updated after solving the energy and species equations, while inputs contain enthalpy, pressure, and Bilger's mixture fraction from the previous iteration. Seven dedicated networks are designed for inferring temperature, density, viscosity, compressibility factor ($\psi = \frac{\partial \rho}{\partial p}$ at constant enthalpy), thermal diffusivity, mixture-averaged diffusion coefficients of each species, and nondimensional standard state enthalpies. Should the variable exhibits inter-species variations, the network is configured with multiple outputs; otherwise, a single-output neural network suffices. Among these variables, the accuracy of temperature, density, and compressibility factor significantly impacts the convergence and stability of CFD simulations.

\subsection{Data generation\label{subsec:data}} \addvspace{10pt}
The training dataset is obtained from the initial 10 time steps of a Taylor-Green Vortex (TGV) simulation. To capture a wide range of operational states during TGV evolution, additional stochastic datasets are generated from the collected results by 
\begin{equation}
    t = c + a_1\frac{max(h)-min(h)}{a_2},
\end{equation}
where $t$ is the training data point, $c$ is the data sampled from CFD, $h$ is the physical variable, and $a_1$, $a_2$ are two random numbers that can be chosen to the fit specific case.
This process requires constraints to prevent non-physical outcomes, ensuring mixture fractions to range from 0 to 1, and densities remain positive. Recognising the multi-scale nature of the broad density spectrum in supercritical conditions, pre-processing techniques are used to optimise the neural network performance. This includes applying the Box-Cox Transformation (BCT) to both density and mixture fraction, highlighting their significant but subtle features, particularly the clustering towards smaller magnitude values.

\subsection{Training\label{subsec:training}}
\addvspace{10pt}
Subsequent to sample data generation, the models are trained using the machine learning frameworks PyTorch and Scikit-learn.The L1 norm is designated as the loss function to minimise the absolute error during training. 
The training of each model is limited to $1,000$ epochs, beyond which no significant improvement in loss reduction is observed.
The wall-clock time required for training each network approximates one hour. This duration is kept short to ensure that the expeditiousness imparted by the deployment of the DNN is not nullified by the training process itself.

\begin{figure}[h!]
\centering
\includegraphics[width=192pt]{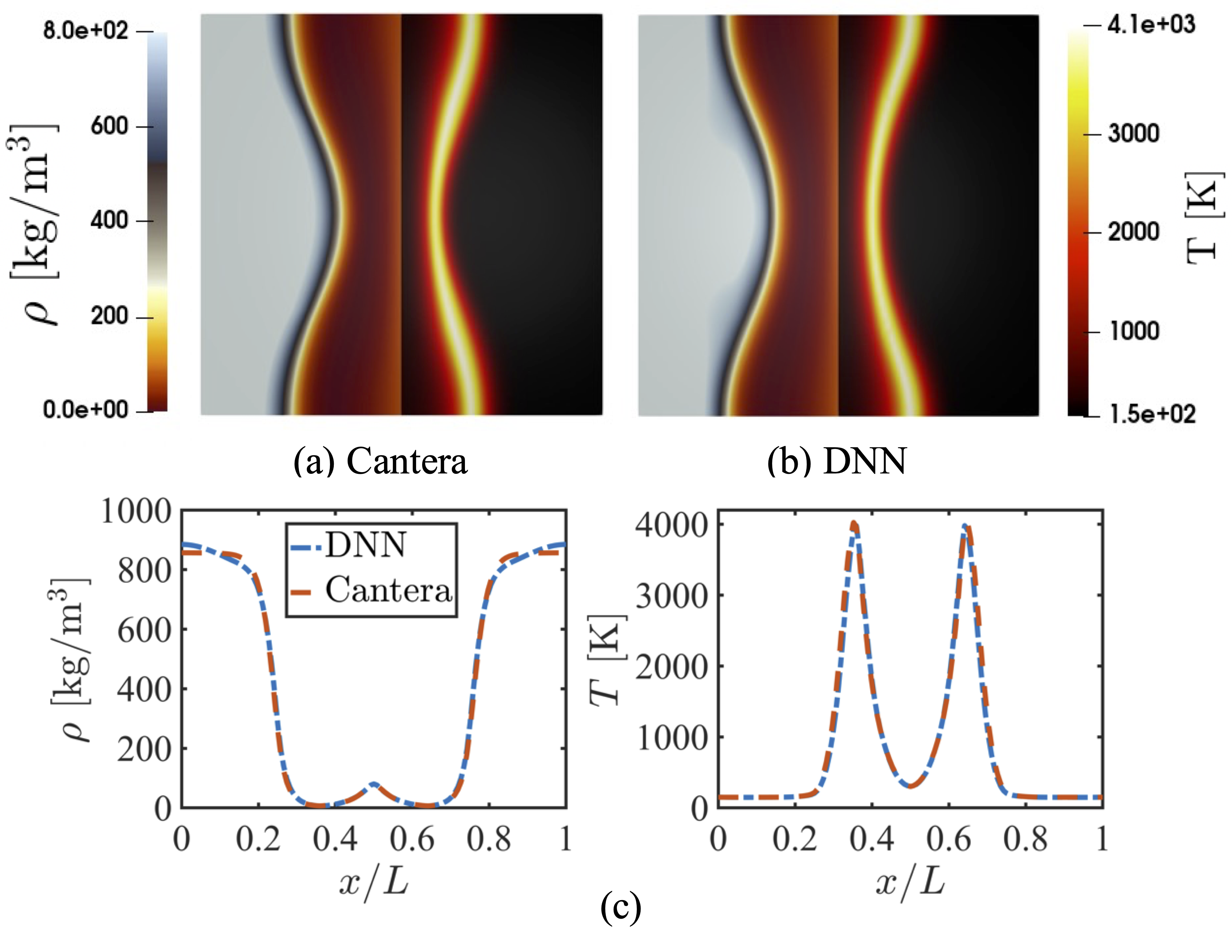}
\caption{\footnotesize The comparison between the results at 0.4 reference time obtained from (a) Cantera and (b) DNN inference for density~(field on the left) and temperature~(field on the right) fields. Values on the central line are plotted in (c).}
\label{dnn_rho}
\end{figure}

\begin{figure}[h!]
\centering
\includegraphics[width=192pt]{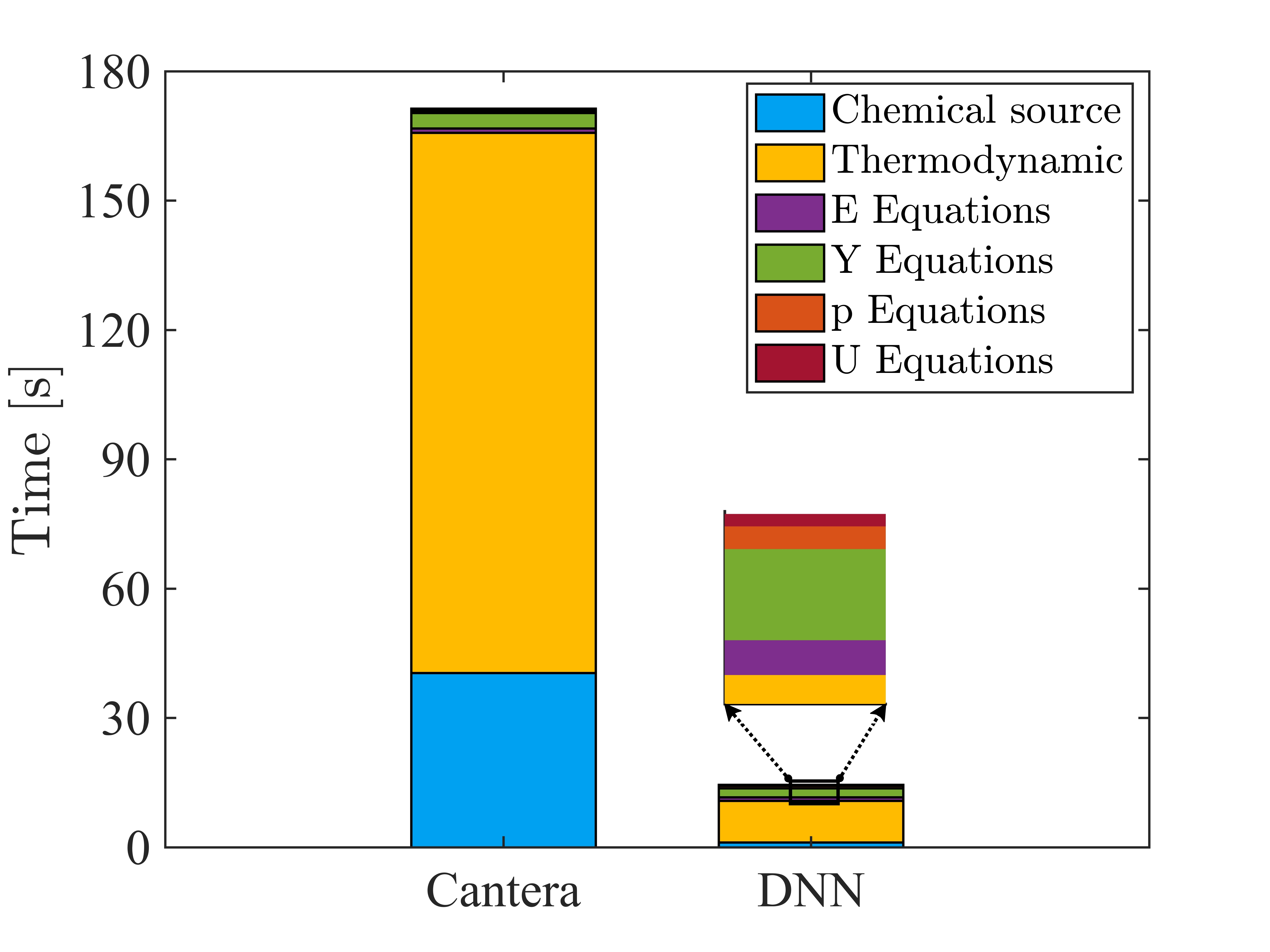}
\caption{\footnotesize Computation time comparison between Cantera and DNN.}
\label{speedup}
\end{figure}

\subsection{Performance\label{subsec:performance}}
\vspace{10 pt}
The fidelity of network predictions is evaluated by \textit{a priori} assessment and comprehensive comparisons with flow field results obtained from CFD simulations. Figure~\ref{dnn} depicts a scatter plot comparing neural network (NN) predictions with labelled data accompanied by a colour-coded gradient to illustrate the distribution of local relative errors. Each plot includes an R2-score, a widely recognised measure of regression model performance, which affirms the strong alignment between NN outputs and the intended targets, suggesting an overall good model accuracy for the thermophysical quantities considered.
Subsequent application of the NN framework to a three-dimensional reacting TGV simulation (discussed in Section~\ref{subsec:3DreactiveTGV}) yields the density $\rho$ and temperature fields presented in Fig.~\ref{dnn_rho}. Both the field and line graphs show satisfying agreement between the DNN and Cantera real fluid package. Employing the relative Mean Squared Error (MSE) defined in~\cite{MILAN2021110567} as a metric for validation, the relative errors in density $\rho$, pressure $p$, and temperature $T$ are confined to $2.1\%$, $1.2\%$, and $9\%$, respectively. This suggests a reasonable predictive accuracy of the NN model for the thee-dimensional supercritical reactive TGV simulation considered in this study.

Finally, computational acceleration provided by the machine learning approach is of primary interest. Figure~\ref{speedup} compares computation time required to advance one time step of the supercritical reactive TGV simulation using Cantera and DNN for thermophysical property calculations with 16 CPU cores. The DNN inference was performed using one additional GPU card. A speed-up factor of 13 is achieved for the thermodynamics when the DNN model is used. A further acceleration strategy is applied for the other time-consuming part $-$ chemical reaction source term integration~\cite{df}, resulting in an overall 12 times speed-up for the entire CFD simulation. 
 
\section{Conclusion \label{conclusion}}
In this work, detailed numerical simulation was performed to study LOX/GCH4 flame-vortex interaction under supercritical conditions. The benchmark configuration of  Taylor-Green Vortex (TGV) interacting with a diffusion flame is modified to accommodate the trans/supercritical LOX/GCH4 reactive mixtures. The results show that under non-reative conditions, the TGV evolution remains similar to that observed in subcritical conditions with slightly more pronounced vorticity generation due to large densities. However, for the reactive case, considerably different flame-vortex interaction behaviours are observed comparing the real gas and ideal gas cases. The large density gradients present in the real gas case result in strong flame stretching and quenching, which pose substantial modelling challenges for this benchmark configuration. Furthermore, a machine learning-based methodology is introduced to lower the computational cost associated with real fluid equation of state and thermodynamic property calculations. The proposed deep neural network (DNN) model shows good predictive capabilities to capture the state variable and transport properties. An impressive speed-up factor of 13 is achieved using the DNN model for the real fluid thermodynamics computation, suggesting an efficient and yet accurate modelling paradigm for future detailed simulation of trans/supercritical reacting flows.

\acknowledgement{Declaration of competing interest} \addvspace{10pt}


The authors declare that they have no known competing financial interests or personal relationships that could have appeared to influence the work reported in this paper.


 \footnotesize
 \baselineskip 9pt


\bibliographystyle{pci}
\bibliography{PCI_LaTeX}


\newpage

\small
\baselineskip 10pt



\end{document}